\documentclass[conference]{IEEEtran}
\pdfoutput=1

\usepackage[dvipsnames]{xcolor}
\usepackage{cite}
\usepackage{soul}
\usepackage{mathtools}
\usepackage{amsmath,amssymb,amsfonts}
\usepackage{comment}
\usepackage{graphicx}
\usepackage{amssymb}
\usepackage{amsmath}
\usepackage{textcomp}
\usepackage[pscoord]{eso-pic}

\usepackage{booktabs}
\usepackage{siunitx}
\usepackage{lipsum}
\usepackage{paralist}
\usepackage{enumitem}
\usepackage{algorithm}
\usepackage[noend]{algpseudocode}
\usepackage[normalem]{ulem}
\usepackage[colorlinks,allcolors=blue,urlcolor=black]{hyperref}
\usepackage{cancel}
\usepackage{balance}
\usepackage{todonotes}

\usepackage{multirow}
\algnewcommand\algorithmicforeach{\textbf{for each}}
\algdef{S}[FOR]{ForEach}[1]{\algorithmicforeach\ #1\ \algorithmicdo}

\def\BibTeX{{\rm B\kern-.05em{\sc i\kern-.025em b}\kern-.08em
    T\kern-.1667em\lower.7ex\hbox{E}\kern-.125emX}}
\def\BibTeX{{\rm B\kern-.05em{\sc i\kern-.025em b}\kern-.08em
    T\kern-.1667em\lower.7ex\hbox{E}\kern-.125emX}}

\usepackage{tikz}
\newcommand*\circled[1]{\tikz[baseline=(char.base)]{
            \node[shape=circle,draw=black, font=\bfseries, fill=white,inner sep=0.5pt,scale=0.85] (char) {\textcolor{black}{#1}};}}

\begin{document}
\bstctlcite{IEEEexample:BSTcontrol}

\title{{\vspace{-1.5cm}\small Accepted for publication at the 40th International Conference On Computer Aided Design (ICCAD 2021). \textcopyright 2021 IEEE.}\vspace{-0.8\baselineskip}
\rule{\textwidth}{0.4pt}\vspace{0.2cm}
Positive/Negative Approximate Multipliers\\ for DNN Accelerators
}

\author{\IEEEauthorblockN{
Ourania Spantidi\IEEEauthorrefmark{2},
Georgios Zervakis\IEEEauthorrefmark{1},
Iraklis Anagnostopoulos\IEEEauthorrefmark{2},
Hussam Amrouch\IEEEauthorrefmark{3},
and J{\"o}rg Henkel\IEEEauthorrefmark{1}
}
\IEEEauthorblockA{\IEEEauthorrefmark{2}School of Electrical, Computer and Biomedical Engineering, Southern Illinois University, Carbondale, U.S.A.}
\IEEEauthorblockA{\IEEEauthorrefmark{3}Chair of Semiconductor Test and Reliability (STAR), University of Stuttgart, Stuttgart, Germany}
\IEEEauthorblockA{\IEEEauthorrefmark{1}Chair for Embedded Systems (CES), Karlsruhe Institute of Technology, Karlsruhe, Germany}
\IEEEauthorblockA{
\IEEEauthorrefmark{2}\{ourania.spantidi,iraklis.anagno\}@siu.edu,
\IEEEauthorrefmark{3}amrouch@iti.uni-stuttgart.de,
\IEEEauthorrefmark{1}\{georgios.zervakis,henkel\}@kit.edu
}\vspace{-20pt}
}

\maketitle

\begin{abstract}

Recent Deep Neural Networks (DNNs) managed to deliver superhuman accuracy levels on many AI tasks. Several applications rely more and more on DNNs to deliver sophisticated services and DNN accelerators are becoming integral components of modern systems-on-chips. DNNs perform millions of arithmetic operations per inference and DNN accelerators integrate thousands of multiply-accumulate units leading to increased energy requirements.  Approximate computing principles are employed to significantly lower the energy consumption of DNN accelerators at the cost of some accuracy loss. Nevertheless, recent research demonstrated that complex DNNs are increasingly sensitive to approximation. 
Hence, the obtained energy savings are often limited when targeting tight accuracy constraints.
In this work, we present a dynamically configurable approximate multiplier that supports three operation modes, i.e., \emph{exact, positive error, and negative error}. In addition, we propose a filter-oriented approximation method to map the weights to the appropriate modes of the approximate multiplier. Our mapping algorithm balances the positive with the negative errors due to the approximate multiplications, aiming at maximizing the energy reduction while minimizing the overall convolution error. We evaluate our approach on multiple DNNs and datasets against state-of-the-art approaches, where our method achieves $18.33\%$ energy gains on average across $7$ NNs on $4$ different datasets for a maximum accuracy drop of only 1\%.

\end{abstract}
\begin{IEEEkeywords}
Approximate Computing, Deep Neural Networks, Multipliers, Low Power
\end{IEEEkeywords}


\section{Introduction}\label{sec:intro}

\IEEEPARstart{W}{ith} the recent and rapid advancements in the area of machine learning, Neural Networks (NNs) have become the driving force for 
embedded devices advancing a variety of domains, such as object detection, speech recognition and more~\cite{jouppi2017datacenter}. However, such devices are generally characterized by limited computing capabilities and they are also operating under strict power budgets  due to tight temperature constraints~\cite{amrouch2020npu}. Furthermore,
NNs, and specifically Deep NNs (DNNs), are continuously evolving, becoming more and more  computationally intensive in order to accommodate the latest research and industry accuracy requirements. During the inference phase of NNs, the dominant arithmetic operation, performed mainly on the convolution and fully connected layers, is the multiply-accumulate (MAC) operation. Accordingly, embedded devices integrate DNN accelerators as a solution to the throughput/latency requirements. Such accelerators comprise large amounts of MAC units, for instance, the Google Edge TPU comprises 4K MAC units~\cite{cass2019taking} and the Samsung embedded NPU integrates 6K MAC units~\cite{park20219}.

Even though the hardware accelerators might be a solution towards addressing the computing limitations of embedded devices, integrating thousands of MAC units in order to keep up with the computational demands, results in increased energy consumption~\cite{amrouch2020npu}. Interestingly, previous research works~\cite{sarwar2018energy,mrazek2019alwann,zervakis2021control,tasoulas2020weight,zervakis2020design,VenkataramaniJPROC2020} have shown that a great amount of these computations can tolerate at least some degree of approximation, thus reducing energy consumption and without sacrificing the NN inference accuracy. Thus, exploiting the principle of approximate computing, we can trade-off the system's energy efficiency with respect to the NN accuracy. This has led to the design and development of approximate circuits, and particularly multipliers~\cite{sarwar2018energy}. As shown in~\cite{VenkataramaniJPROC2020}, approximately 90\% of the DNN computations are general matrix multiplication operations. Hence, employing approximate multipliers delivers considerable energy reduction with respect to the entire accelerator.

Current techniques for designing approximate multipliers, mainly focus on the introduction of low error applying also retraining to recover the accuracy loss~\cite{sarwar2018energy}. However, this is not always feasible due to proprietary datasets and models~\cite{mrazek2019alwann,zervakis2021control}. Moreover, retraining requires to simulate the approximate multiplier, precluding the exploitation of hardware optimizations (e.g., AVX instructions) and exponentially increases the training time, especially for deeper networks~\cite{mrazek2019alwann}.

\textbf{In this work}, we target DNN inference and apply approximation to maximize energy gains without significant losses in accuracy. Particularly,  
we present a \textit{reconfigurable multiplier, that follows a different approach than state-of-the-art, by comprising one exact mode of operation and two synergistic approximate modes, one positive and one negative, that aim to balance the introduced error.}
Additionally, we present an approximation strategy that assigns NN operations to specific approximation modes based on the respective layer, filter, and weight value of the operation.
The proposed approach doesn't require retraining and aims to minimize the convolution error by introducing, in a directed manner, positive and negative approximation in the performed multiplications.
The contributions of our work are summarized in the following points:\\
\begin{inparaenum}[(1)]
\item We design an approximate multiplier that approximates the generation of the partial products and comprises three operation modes: the Zero Error (ZE) mode (i.e., exact), the Positive Error (PE) mode, and the Negative Error (NE) mode.\\
\item By approximating more or less partial products, we control the applied approximation and tune the accuracy-energy reduction trade-off as required.\\
\item We present a methodology where for each NN filter, we exploit the two synergistic modes of the presented multiplier, and group all the weight values into two balanced summation sets, with the final goal being the convolution error converging to zero and consequently achieving high inference accuracy.
\end{inparaenum}

\section{Related works}\label{sec:Related}

Approximate DNN accelerators have attracted a vast research interest.
A significant fraction of a DNN operations (about 90\%) is attributed to GEMM operations, i.e., convolution and matrix multiplications~\cite{VenkataramaniJPROC2020}.
Such computations rely on MAC units and the state of the art approximates the multipliers to boost the energy efficiency of the overall accelerator.
Cartesian Genetic Programming is used in~\cite{mrazek2016design,ansari2019improving} to generate fixed approximate multipliers and replace the accurate ones, achieving high hardware gains for a minimal accuracy loss or even accuracy improvement.
\cite{sarwar2018energy} introduced a multiplier-less artificial neuron that is based on additions only.
Nevertheless, \cite{mrazek2016design,sarwar2018energy,ansari2019improving} require retraining, which as aforementioned is not always feasible.

In~\cite{mrazek2019alwann} the authors avoid retraining and use layer-based approximation
in which a different fixed approximate multiplier from~\cite{mrazek2017evoapproxsb} is used per layer.
In addition, a weight tuning method is employed targeting to reduce the introduced multiplication error.
The work in~\cite{mrazek2020using} extends the approximate multipliers of~\cite{mrazek2017evoapproxsb} and shows that, in simple DNNs, high energy gains and minimal accuracy loss can be achieved, even without retraining.
However, for more complex DNNs the energy gains are not maintained.
Acknowledging the need for runtime reconfiguration, \cite{zervakis2020design} generates approximate multipliers with dynamically reconfigurable accuracy and uses them to apply layer-wise approximation in DNNs by changing the multiplier's accuracy mode per layer.
The work in~\cite{tasoulas2020weight} uses~\cite{zervakis2020design} to generate low variance approximate reconfigurable multipliers, and proposes a weight-oriented approximation for DNN inference.
\cite{hanif2019cann} employs a curable approximation in which the MAC's adder is split into low and high parts and the carry of low part is accumulated by the neighboring MAC unit.
The carry of the last MAC unit is not corrected.
However, \cite{hanif2019cann} is evaluated on the LeNet architecture, a very shallow architecture which cannot provide the amount of operations recent DNNs do.
The work in~\cite{riaz2020caxcnn} uses the  Canonic Sign Digit Encoding to represent the weights and employs truncation as an approximation method.
The architecture proposed in~\cite{hammad9cnn} uses Dynamic and Static Segmented approximate multipliers that support high and low precision for the size of the segment.
A trainable input classifier is used to select the required precision per segment.
\cite{park2021design} targets energy consumption minimization of MAC-based signal processing algorithms.
\cite{park2021design} utilizes different fixed approximate multipliers in an interleaved manner to compensate errors during the accumulate operations.
Nevertheless,~\cite{hammad9cnn,park2021design} consider $16$-bit inference and can be deemed inapplicable in modern DNN accelerators that use mainly $8$-bit precision~\cite{jouppi2017datacenter}.
Finally,~\cite{zervakis2021control} introduces a control variate approximation to recover the accuracy loss due to approximate multiplications.
The overall convolution error is estimated at runtime and it is finally accumulated in the output result. However, for the error accumulation, it requires an additional column of MAC units.

\textbf{Our work differentiates} from the state of the art as since it does not require retraining and it employs a reconfigurable multiplier that supports positive/negative error as well as accurate execution. 
In addition, we propose a filter-oriented weight mapping methodology to map DNNs to the modes of the approximate multiplier so that given accuracy loss constraints are satisfied.

\section{Proposed Methodology}\label{sec:Methodology}
\begin{figure}
    \centering
    \includegraphics[width=\columnwidth]{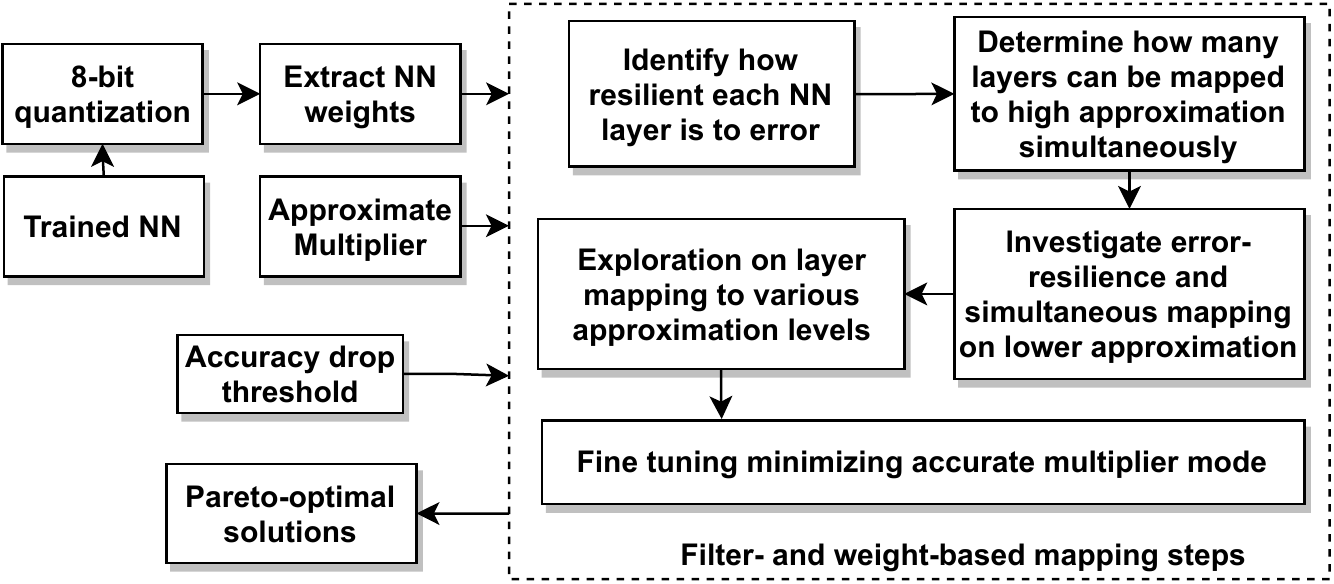}
    \caption{An overview of the proposed methodology}
    \label{fig:methodology_overview}
    \vspace{-15pt}
\end{figure}

An overview of our approach is depicted in Fig.~\ref{fig:methodology_overview}.
First, we present our positive/negative approximate multiplier and show a rigorous error analysis that is exploited in the error optimization of our mapping methodology (Section~\ref{sec:perfmult}). Then, given a trained DNN, we quantize weights and activations to 8-bit (in the range [0, 255]~\cite{jacob2018quantization}) and we apply our mapping methodology, responsible for assigning the weights to the modes of our multiplier (Section~\ref{sec:mapping}).

\subsection{Positive/Negative  Approximate  Multiplier in NNs}\label{sec:perfmult}
The most complex computation in the CNN inference phase is the convolution operation. The latter is expressed as:
\begin{equation}\label{eq:conv}
G=B+\sum_{i=1}^{k}{W_i\cdot A_i},
\end{equation}
where $W_i$ are the weights, $A_i$ are the input activations, and $B$ is the bias of the neuron.
We assume a microarchitecture similar to Google TPU that comprises a big
systolic MAC array~\cite{jouppi2017datacenter,cass2019taking}.
In addition, we consider a weight-stationary mapping and we replace the exact multipliers with approximate ones. 
Denoting $\epsilon_i$ the error of the approximate multiplication:
\begin{equation}\label{eq:multerror}
\epsilon_i = W_i\cdot A_i -  W_i\cdot A_i|_{approx}
\end{equation}
the error $\epsilon_G $ of the approximate convolution is given by:
\begin{equation}\label{eq:converror}
\epsilon_G = G - G|_{approx} = \sum_{i=1}^{k}{\epsilon_i}.
\end{equation}

In~\cite{ZervakisTVLSI2016}, the authors proposed an approximate multiplier with predictable (known a priori) error.
The multiplier of~\cite{ZervakisTVLSI2016} eliminates the generation of some partial products (they are set to zero) and thus less partial products need to be accumulated.
The technique in \cite{ZervakisTVLSI2016} always leads to positive error as the approximate product is always smaller than the exact one.
Consequently, when approximating (eliminating) the $z$ least partial products, the error $e_i$ is given by:
\begin{equation}\label{eq:peerr}
\begin{split}
\epsilon_i  &= W_i\cdot A_i - W_i\cdot (A_i-A_i\text{ mod }2^z) \\
            &= W_i\cdot r_i, \text{ with } r_i = A_i\text{ mod }2^z.
\end{split}
\end{equation}

Hence, the average multiplication error and the error variance $\forall A_i$ of~\cite{ZervakisTVLSI2016} are given by:
\begin{equation}\label{eq:pestats}
\begin{split}
\mathrm{E}[\epsilon_i] &=\frac{2^z-1}{2}W_i\\
\mathrm{Var}(\epsilon_i) &=\frac{2^{2z}-1}{12}W_i.
\end{split}
\end{equation}

\begin{figure}
    \centering
    \includegraphics[width=\columnwidth]{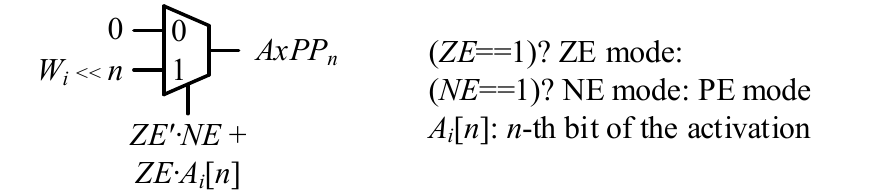}
    \caption{Run-time approximate generation of the \textit{n}-th partial product. ZE, PE, and NE modes are supported. Simple partial product generation is used as an illustrative example.}
    \label{fig:mux}
\end{figure}

The authors in~\cite{LeonMICRO2018} proposed to use switches and control, at run-time, the number of partial products that will be approximated (i.e., set the $z$ value at run-time).
Hence,~\cite{LeonMICRO2018} supports also exact multiplications (i.e., when $z=0$).
In our work, we extend~\cite{LeonMICRO2018} to support three different modes: \emph{Zero Error (ZE), Positive Error (PE), and Negative Error (NE)}.
The ZE mode refers to the exact operation, in which no error is introduced in the multiplication.
In the PE mode, the $z$ least partial products are perforated and thus positive error is obtained.
In the NE mode, we force the generation and accumulation of the $z$ least partial products and thus negative error is obtained.
Fig.~\ref{fig:mux} presents how the three operating modes (ZE, PE, and NE) can be configured at run-time.
Considering the weight-stationary mapping, in both NE and PE modes, the $z$ partial products remain fixed at run-time (for several cycles)
leading to reduced switching activity and thus high power gains.

Since, in the NE mode, we force the generation of the partial products, the multiplication error $\epsilon_i$ is given by:
\begin{equation}\label{eq:neerr}
\begin{split}
\epsilon_i  &= W_i\cdot A_i - W_i\cdot \big((A_i+(2^z-1-A_i\text{ mod }2^z)\big)\\
            &= -W_i\cdot(2^z-r_i-1), \text{ with } r_i = A_i\text{ mod }2^z.
\end{split}
\end{equation}
Thus, in the NE mode, the average multiplication error and the error variance  $\forall A_i$ are given by:
\begin{equation}\label{eq:nestats}
\begin{split}
\mathrm{E}[\epsilon_i] &=-\frac{2^z-1}{2}W_i\\
\mathrm{Var}(\epsilon_i) &=\frac{2^{2z}-1}{12}W_i.
\end{split}
\end{equation}

As a result, from~\eqref{eq:pestats} and~\eqref{eq:nestats}, the average error and the error variance, $\forall A_i$, of our approximate multiplier are given by:
\begin{equation}\label{eq:axmult}
\begin{split}
\mathrm{E}[\epsilon_i] &=s\frac{2^z-1}{2}W_i\\
\mathrm{Var}(\epsilon_i) &=\frac{2^{2z}-1}{12}W_i,
\end{split}
\end{equation}
where $s=1$ in the PE mode, $s=-1$ in the NE mode, and $z=0$ in the ZE mode.

Without any loss of generality, each multiplier in the DNN accelerator can be configured to a different mode, i.e., each multiplier features different $s$ and $z$ values, named $s_i$ and $z_i$ respectively.
Therefore, using~\eqref{eq:converror} and~\eqref{eq:axmult}, the average convolution error $\mathrm{E}[\epsilon_G]$ is given by:
\begin{equation}\label{eq:avgconv}
\mathrm{E}[\epsilon_G] = \sum_{i=1}^{k}{\mathrm{E}[\epsilon_i]}  = \sum_{i=1}^{k}{s_i\frac{2^{z_i}-1}{2}W_i}
\end{equation}
and the convolution error variance $\mathrm{Var}(\epsilon_{G})$ is given by:
 \begin{equation}\label{eq:varcov}
\begin{split}
\mathrm{Var}(\epsilon_{G}) & = \sum_{i=1}^{k}{\mathrm{Var}(\epsilon_i)}+\sum_{\mathclap{1\leq i \leq j \leq k}}{\mathrm{Cov}(\epsilon_i, \epsilon_j)} \\
& = \sum_{i=1}^{k}{\frac{2^{2z_i}-1}{12}W_i} + \sum_{\mathclap{1\leq i \leq j \leq k}}{W_iW_j\cancelto{0}{\mathrm{Cov}(r_i,r_j)}} \\
& = \sum_{i=1}^{k}{\frac{2^{2z_i}-1}{12}W_i}.
\end{split}
\end{equation}\noindent
in which $r_i$ and $r_j$ are uncorrelated and thus their covariance $\mathrm{Cov}(r_i,r_j)$ is zero.

Exploiting that \eqref{eq:avgconv} and \eqref{eq:varcov} depend only on the weights and leveraging the fact that the weight values are obtained after training and quantization, 
we can minimize the convolution error (i.e., minimize~\eqref{eq:avgconv} and \eqref{eq:varcov}) by carefully setting the approximation parameters of each multiplier (i.e., $s_i$ and $z_i$).

Finally, Table~\ref{tab:energy} shows the achieved energy reduction of the PE and NE modes for different $z$ values.
As it can be seen, the energy gains increase as the value of $z$ increases.
However, the magnitude of the multiplication error, both in PE and in NE, becomes larger as well, as calculated by~\eqref{eq:axmult}.
Therefore, in Section~\ref{sec:mapping} we present a method to map the weights to specific modes in order to keep the overall inference accuracy loss low.
The experimental setup and tool flow used to obtain the values reported in Table~\ref{tab:energy} are described in Section~\ref{sec:Evaluations}.

\begin{table}[]
\centering
\caption{Energy gains of our $8$-bit Positive/Negative Approximate Multiplier}
\label{tab:energy}
\begin{tabular}{c|ccc}
Mode & $z=1$ & $z=2$ & $z=3$ \\
\toprule
\toprule
PE   &  8.3\%  & 20.23\%  & 36.6\%    \\
NE   &  5.5\%  & 16.17\%  & 31.8\% \\
\bottomrule
\end{tabular}
\end{table}

\subsection{Filter- and weight-based mapping}\label{sec:mapping}

\begin{figure}
    \centering
    \resizebox{0.93\columnwidth}{!}{\includegraphics{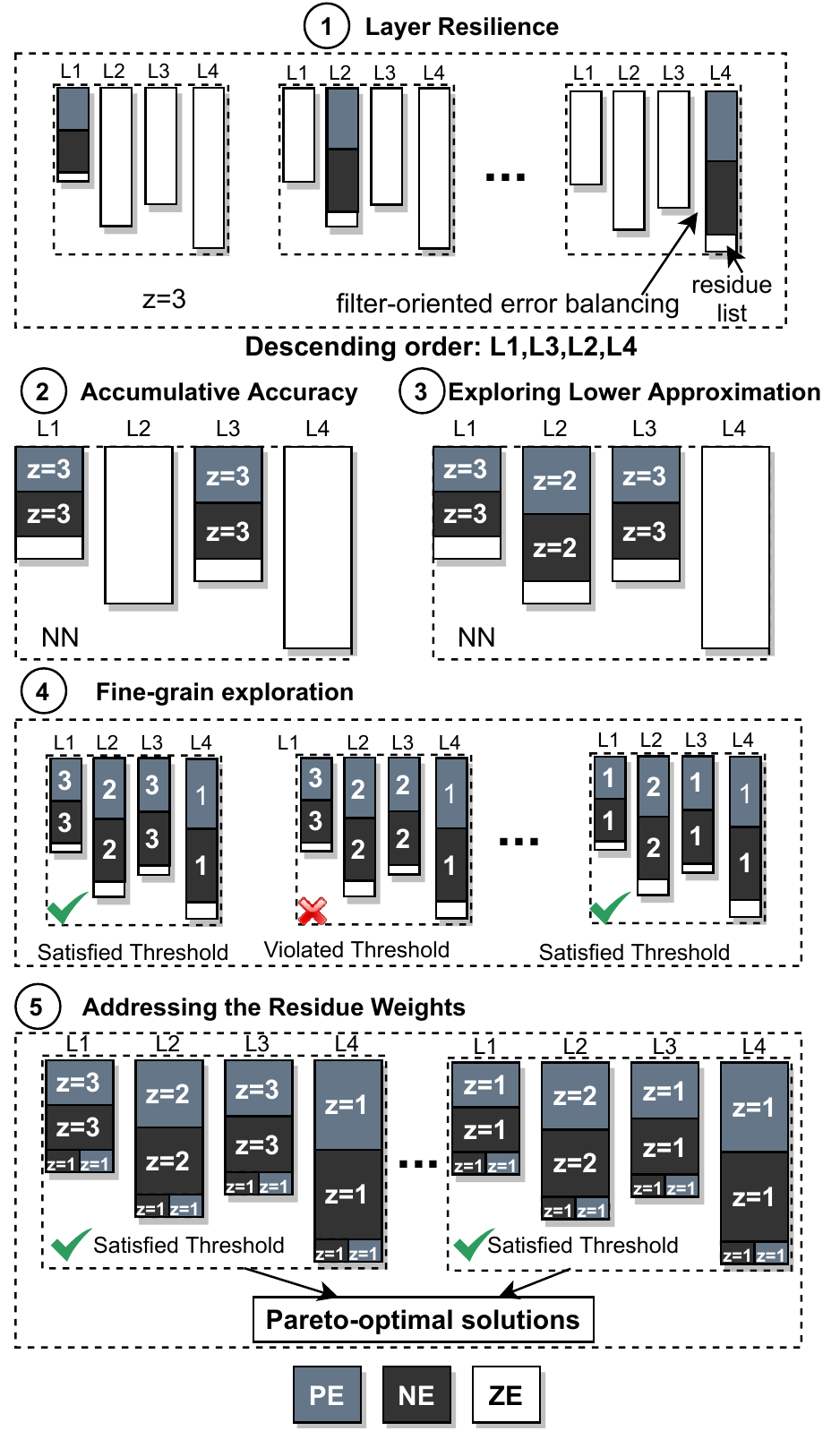}}
    \caption{Illustrative example of the proposed five-step methodology.}
    \label{fig:methodology_eg}
\end{figure}

In this section, we present a filter-oriented method for mapping the weights of an NN to the three aforementioned modes of the approximate multiplier as well as deciding the value of $z$ for each one of them.
For our analysis, the available values for $z$ are $1$, $2$, and $3$. We also tested values greater than $3$, but the introduced error was very large and violated our tight accuracy thresholds. Our five-step mapping procedure aims to satisfy a given accuracy drop threshold while maximizing the number of weights that are assigned to high $z$ values. Fig.~\ref{fig:methodology_eg} depicts an illustrative example of the steps.

\underline{\textit{Step 1 - Layer resilience:}} The goal of this step is to identify how error resilient each convolution layer of the targeted NN is (Fig~\ref{fig:methodology_eg}~\circled{1}). Initially, we consider that all weights are assigned to the exact mode (i.e., ZE mode). Then, for each layer of the NN separately, starting from the first one, we count the occurrences of each weight value \emph{per filter}.
We define as $w_{i,f,l}$ the number of times that weight $i$ occurs in layer $l$ of filter $f$.
We take advantage of the positive/negative architecture of the proposed multiplier and we assign the $w_{i,f,l}/2$ weights to the PE mode and the rest half of the weights to the NE mode in order to cancel out the introduced error (see \eqref{eq:avgconv}).
If $w_{i,f,l} \% 2 \neq 0$, we map $\left \lfloor{w_{i,f,l}/2}\right \rfloor$ times the weight $i$ to the PE and NE modes and the remaining occurrence of $i$ to the ZE mode, keeping it also in a residue list, unique for each filter, to be used in the last step for further tuning. We call this concept \emph{filter-oriented error balancing.} 
For this particular step, we set $z=3$ (for all weights) for the PE and NE modes, as it introduces the highest error compared to $z=2$ and $z=1$, achieving higher energy gains.
Since the procedure described above is performed for each layer separately, we record the accuracy output of the network each time and we determine which layers are more sensitive to approximate multiplications, and which layers show small or no drop in the final network accuracy. Once we have obtained the layer resilience information, the output of this step is a list of convolution layers of the network sorted based on error resiliency (i.e., highest accuracy to lowest inference accuracy).
At this point, although the weights assigned to the PE and NE modes (positive and negative error) are balanced, the convolution error is still defined by its variance, as \eqref{eq:varcov} shows.
The latter depends exponentially on $z$.
Hence, this steps sorts the layers with respect to $z$ tolerance.

\underline{\textit{Step 2 - Accumulative inference accuracy:}} In this step, our goal is to find how many layers can be mapped to high approximation (high $z$) simultaneously, using the filter-oriented error balancing method presented in Step 1 (Fig~\ref{fig:methodology_eg}~\circled{2}). Thus, starting from the most error resilient layer of the network towards the least resilient one, we map the convolution layers to the PE and NE modes following the procedure described previously, but this time in an accumulative way, still using $z=3$. We stop this step once we have reached the accuracy drop threshold.

\underline{\textit{Step 3 - Exploring lower approximation:}} 
In this step, we repeat the procedure that was described in Steps 1 and 2, however in this case we set $z=2$ (Fig~\ref{fig:methodology_eg}~\circled{3}). However, we do not perform the layer resilience and accumulative accuracy process to all convolution layers of the network, but only to the remaining ones out of the procedure described in Step 2. When this step is finished, in most cases we are left with portion of the network's convolution layers mapped to PE/NE using $z=3$, some mapped using $z=2$, and the remainder of the layers mapped to ZE.

\underline{\textit{Step 4 - Fine-grain exploration:}} 
Up to this point, the actions described in Steps 1-3 let us reduce the search space. Particularly, we explored the error resilience of the NN layers for $z=3$ and $z=2$ but there are still layers entirely mapped to the ZE mode. These layers, that cannot tolerate approximation for $z=3$ and $z=2$, will be mapped to the PE/NE modes with $z=1$. However, this new mapping to $z=1$ can severely impact NN accuracy, violating the accuracy threshold. Hence, in this step we perform a fine-grain exploration for different mapping combinations among the different $z$ values in order to balance this newly introduced error (Fig~\ref{fig:methodology_eg}~\circled{4}). Additionally, this step lets us perform a more thorough search for more valid mappings and let us create a Pareto-front. 
The exploration is performed in three parts.
First, we start moving one by one all the layers mapped to $z=3$ to $z=2$, starting from the layer that was the last one to be mapped to $z=3$ and we keep the solutions that satisfy the accuracy drop threshold requirement. 
Second, we follow the same concept for the layers mapped to approximate modes with $z=2$, this time moving them to $z=1$, while still keeping each mapping combination that satisfies the accuracy drop threshold. This part is a step towards maximizing energy savings, since all the layers mapped to $z=3$ remain this way, but the layers mapped to $z=2$ are being moved to $z=1$ in order to reduce the accuracy drop.
Finally, all layers initially mapped to PE/NE with $z=3$ are moved to $z=1$ approximate modes, while keeping each mapping combination along the way that does not violate the accuracy threshold requirement. This is another way to drastically reduce the introduced error, as the approximation under $z=3$ is more aggressive, mostly relying on the layers mapped to $z=2$ for energy gains. Overall, the output of this step is a list of valid mappings, with varying energy savings, utilized in Step 5 for final tuning. 

\underline{\textit{Step 5 - Addressing the residue weights:}} 
So far, in all previous steps, the weights included into the residue lists of each filter, described in Step 1, are being mapped to the ZE mode. Thus, in this step we use all the mapping configurations found so far, that satisfy the accuracy threshold, and we map all these residue weights to either the PE or NE mode (Fig~\ref{fig:methodology_eg}~\circled{5}). Hence, for each filter we partition the residue list into \emph{two balanced summation sets} using the Largest Differencing Method~\cite{karmarkar1982efficient} (LDM) algorithm. Then, we map all weight values in the first set to the PE mode, and the weight values in the second set to the NE mode. Again, in this step we keep all the solutions that satisfy the accuracy requirement. For all the solutions, the residue weights will be mapped to approximate modes starting with $z=1$, then $z=2$ and finally $z=3$, in an attempt to push the approximation for better energy results.

Overall, targeting high energy gains, our mapping methodology aims in assigning each weight to either PE or NE with a high $z$ value (see Table~\ref{tab:energy}).
Steps 1-4 perform an exploration in which entire layers are approximated (see mapping procedure in Step 1) using a greedy procedure that tries to find the highest $z$ value per layer.
After Step 4, the focus is given on the residue weights, which up to that point are mapped to ZE.
Considering~\eqref{eq:avgconv} and that up to now the positive and negative error weights are completely balanced, the average convolution error in steps 1-4 is zero.
Therefore, only the convolution error variance~\eqref{eq:varcov} affects the accuracy.
Finally, in step 5 we focus on assigning residue weights to non-ZE modes (i.e., boost further the energy gains).
Note that, LDM aims to create subsets whose sums are as equal as possible, but it does not guarantee a balanced final partitioning.
Thus, after step 5, \eqref{eq:avgconv} is close to zero (as we discuss later) but might not be actual zero.
For this reason also, applying LDM from the beginning (Step 1) would lead to sub-optimal solutions.
Using LDM for all filter weights, would result to a biased error (non zero~\eqref{eq:avgconv}) and thus, during the $z$ optimization both \eqref{eq:avgconv} and~\eqref{eq:varcov} would contribute to the accuracy loss, resulting in smaller $z$ values per layer and/or more complex $z$ allocation procedure.

\begin{figure}
    \centering
    \includegraphics[width=0.9\columnwidth]{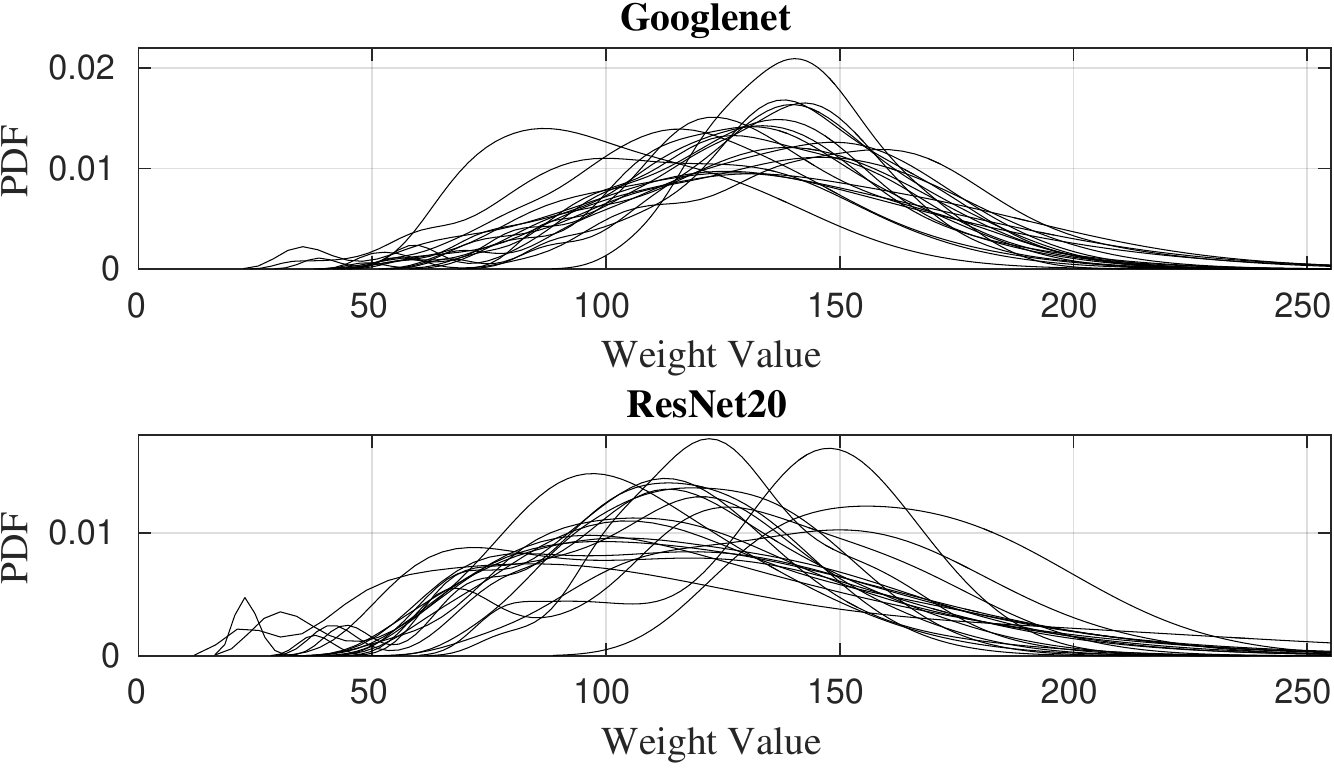}
    \caption{Distributions of weight values for all layers of the GoogleNet and ResNet20 NNs on the CIFAR-10 dataset under 8-bit quantization.}
    \label{fig:weight_act_distr}
\end{figure}

Considering \eqref{eq:avgconv}, the efficiency of the error balancing (i.e., how close \eqref{eq:avgconv} will be to zero) depends on the weight values.
Weight values close to each other increase the probability of error cancellation when employing our positive/negative approximation.
Fig.~\ref{fig:weight_act_distr} shows the weight value distributions for two different NNs: GoogleNet~\cite{szegedy2015going} and ResNet20~\cite{he2016deep} on the CIFAR-10 dataset.
As shown, for both NNs, the weight distributions are close to normal and weight values feature low dispersion.
Finally, setting the mode of operation is seamlessly performed at run-time as the mapping decision is stored with the weights values (i.e., $3$ bits per weight to encode $z$, ZE, and NE). As described in~\cite{riaz2020caxcnn}, targeting recent batch processing DNN accelerators, the storage requirements for similar methods are low since the required memory space is averaged over the entire batch.

\section{Results and Evaluation}\label{sec:Evaluations}

In this section, we provide the experimental evaluation of our proposed method in terms of energy savings and accuracy loss. 
As MAC operations consume a very significant portion of total energy cost, we evaluate the energy reduction w.r.t. the MAC operations.
Note that MAC units are the basic building block of any DNN accelerator.
Additionally, we present comparative results against a variety of state-of-the-art techniques.
For the accuracy evaluation we consider seven DNNs of varying size and characteristics: ResNet20~\cite{he2016deep}, ResNet32~\cite{he2016deep}, ResNet44~\cite{he2016deep}, ResNet56~\cite{he2016deep}, MobileNetv2~\cite{sandler2018mobilenetv2}, GoogleNet~\cite{szegedy2015going}, and ShuffleNet~\cite{zhang2018shufflenet}.
The DNNs were trained on four different datasets: CIFAR-10~\cite{krizhevsky2009learning}, CIFAR-100~\cite{krizhevsky2009learning}, GTSRB~\cite{stallkamp2012man} and LISA~\cite{mogelmose2012vision}.
Overall, $28$ models are considered in our analysis.
In all experiments, 8-bit post-training quantization is used~\cite{jacob2018quantization}. 

\subsection{Overview of Methods under Comparison }\label{sec:methods}

To evaluate our method, we conducted evaluation experiments with other state-of-the-art methods that employ approximate computing techniques, such as fixed approximation across all layers of an NN, or similar fine-grain weight-based approximation mapping. Specifically, we chose the following methods for evaluation comparison:

\underline{Exact}: This method uses exact 8-bit multipliers and is therefore the baseline for our experiments.

\underline{ALWANN}~\cite{mrazek2019alwann}: This method utilizes approximate multipliers from the library in~\cite{mrazek2017evoapproxsb} and employs weight-tuning to minimize the error that the approximate multiplications incur. Note that all multipliers used in this method are fixed and do not comprise different modes of operation. Additionally, this method utilizes non-uniform approximate architectures across the network (i.e., different approximate multiplier per layer) eliminating flexibility and applicability to other networks and datasets when implemented in hardware. For this reason and for fair comparisons, we consider an homogeneous architecture for~\cite{mrazek2019alwann}. In our evaluation, for each use case, we considered all of the Pareto-optimal approximate multipliers described in~\cite{mrazek2017evoapproxsb}, as different NN might require different approximate multiplier from~\cite{mrazek2017evoapproxsb} to satisfy the accuracy loss threshold.

\underline{LVRM}~\cite{tasoulas2020weight}: This is a more fine-grain weight mapping approach that employs a low-variance reconfigurable multiplier and additionally applies a constant error correction term by modifying the biases of the filters.

\underline{ConVar}~\cite{zervakis2021control}: This work uses fix approximation enhanced with a run-time error correction method. \cite{zervakis2021control} induces high approximation at multiplier level to achieve high energy gains and relies on the error correction to achieve high accuracy at convolution and consequently inference levels.

\underline{Filter Balanced Sets (FBS)}: In this method we use the proposed positive/negative multiplier and we employ the concept of LDM on all the weights of all the layers to create two balanced summation sets per filter, instead of applying this concept on just the residue weights as we do in our methodology. By comparing with this method, we want to showcase that just creating balanced sets per layer (from step 1) leads to a biased error and suboptimal results. For fair comparison, we tried all $z$ combinations and we selected the one that satisfies the accuracy thresholds and yields the highest energy gains.

All the aforementioned methods do not require retraining as our  methodology.
In addition, they enable us to evaluate our work against the state of the art, i.e., fixed approximation with statistical error correction~\cite{mrazek2019alwann} and~\cite{zervakis2021control}, but also against more fine-grain run-time reconfigurable approximation~\cite{tasoulas2020weight}.

\emph{We additionally evaluated the methods presented in~\cite{sarwar2018energy,ansari2019improving,mrazek2020using} and~\cite{hammad9cnn}}, however they are not included in our analysis for the following reasons. Both works in~\cite{sarwar2018energy,ansari2019improving} require the retraining of the considered NNs. On the other hand, our proposed method, ALWANN~\cite{mrazek2019alwann}, LVRM~\cite{tasoulas2020weight}, and ConVar~\cite{zervakis2021control} do not require retraining eliminating also the associated time overhead. By bypassing NN retraining, the accuracy delivered by~\cite{sarwar2018energy,ansari2019improving} was poor and did not satisfy any of the considered accuracy thresholds. Furthermore, the work in~\cite{hammad9cnn} is based on 16-bit inference and also produced very poor accuracy when considering 8-bit quantization, as in our analysis. Additionally, although the work in~\cite{mrazek2020using} provided acceptable results for the CIFAR-10 dataset, it did not result in admissible accuracy losses for the CIFAR-100 dataset. The latter is in compliance with the authors conclusion that for simple models the retraining can be avoided when using approximate multipliers, but for complex ones the accuracy drops significantly without retraining. Consequently, we did not include the aforementioned works in our evaluation, since we aim for strict accuracy loss constraints that, almost always~\cite{sarwar2018energy,ansari2019improving,mrazek2020using,hammad9cnn} failed to satisfy.

\subsection{Experimental Setup}

As we target high accuracy, we consider the following accuracy drop thresholds: $0.5\%$, $0.75\%$, and $1\%$. All the aforementioned NNs are trained on each dataset described above. 
Specifically, all NNs are trained using the Tensorflow machine learning library~\cite{tensorflow2015-whitepaper}, and are then frozen and quantized to 8-bit. The accuracy evaluations are performed by describing in \texttt{C} all the approximate multipliers and using the approximate extension of Tensorflow proposed in~\cite{VaverkaDATE2020}.
Accuracy loss is calculated w.r.t. the accuracy achieved by the 8-bit quantized model with exact multiplications. 

Regarding the energy gains, we describe all the examined MAC units in Verilog RTL and industry-strength tools are used for the hardware analysis.
All the MAC units are synthesized using Synopsys Design Compiler and are mapped to a $14$nm technology library calibrated with Intel data~\cite{amrouch2020npu}.
The \texttt{compile\_ultra} command is used for synthesis targeting the maximum frequency that the exact design achieves.
We run post-synthesis timing simulations using Mentor Questasim and $1$ million randomly generated inputs to capture the switching activity of the MAC units.
The switching activity is fed to Synopsys PrimeTime to calculate the power consumption.
In each MAC unit we replace the multiplier with the respective approximate one.
To be in compliance with~\cite{mrazek2019alwann} and~\cite{tasoulas2020weight}, in order to implement our approximate multiplier we used the exact multiplier (1JFF) from~\cite{mrazek2017evoapproxsb} as baseline.
Similarly, 1JFF is used in the exact MAC unit.
\begin{figure*}
    \centering
    \includegraphics[width=0.95\textwidth]{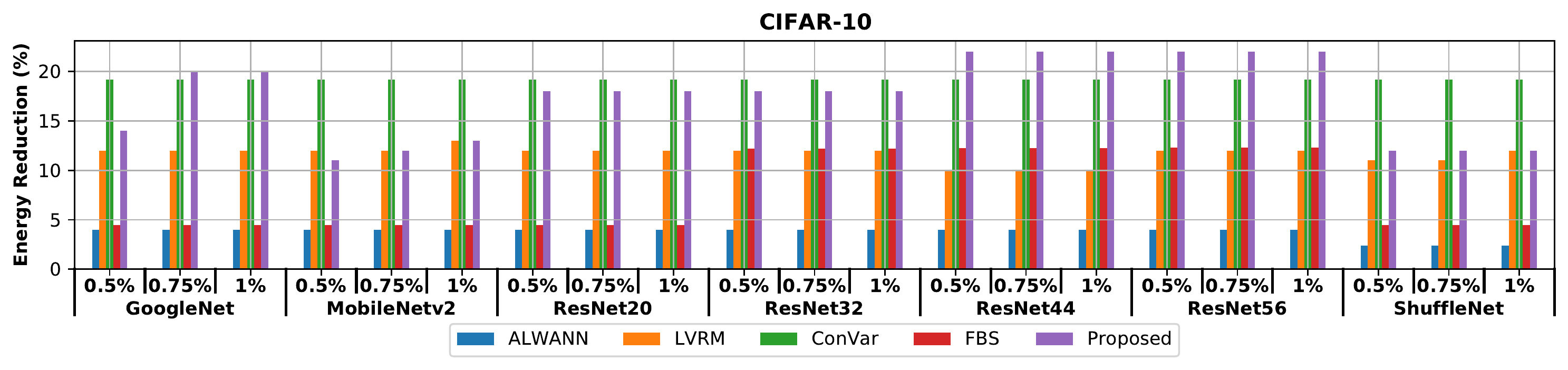}
    \caption{CIFAR-10 normalized MAC operation energy savings.}
    \label{fig:cifar10_results}
\end{figure*}

\begin{figure*}
    \centering
    \includegraphics[width=0.95\textwidth]{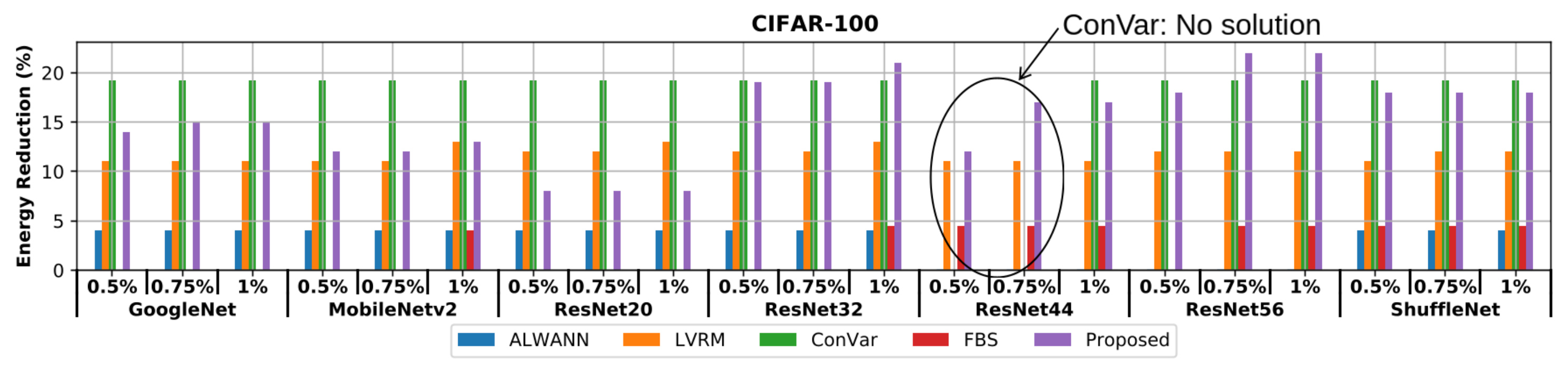}
    \caption{CIFAR-100 normalized MAC operation energy savings. For the $0.5\%$ and $0.75\%$ thresholds for the ResNet44 case, ConVar~\cite{zervakis2021control} could not produce acceptable solutions, while ALWANN~\cite{mrazek2019alwann} resulted in nearly $0\%$ gains in energy for ResNet44 and ResNet56.}
    \label{fig:cifar100_results}
\end{figure*}

\subsection{Results}

\begin{figure*}
    \centering
    \includegraphics[width=0.95\textwidth]{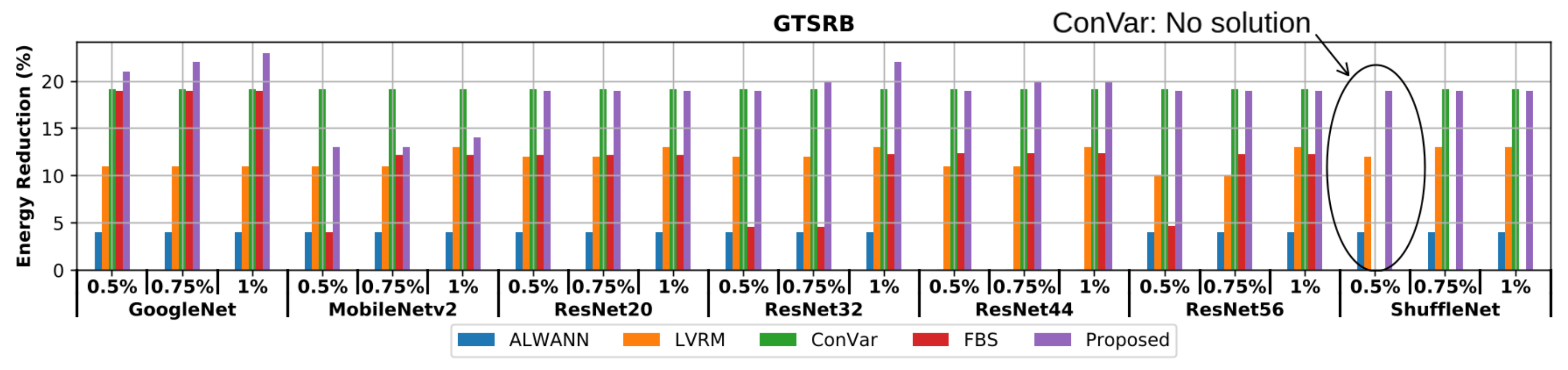}
    \caption{GTSRB normalized MAC operation energy savings. For the $0.5\%$ threshold for the ShuffleNet case, ConVar~\cite{zervakis2021control} could not produce acceptable solutions, while ALWANN~\cite{mrazek2019alwann} resulted in nearly $0\%$ gains in energy for ResNet44.}
    \label{fig:gtsrb_results}
\end{figure*}

\begin{figure*}
    \centering
    \includegraphics[width=0.95\textwidth]{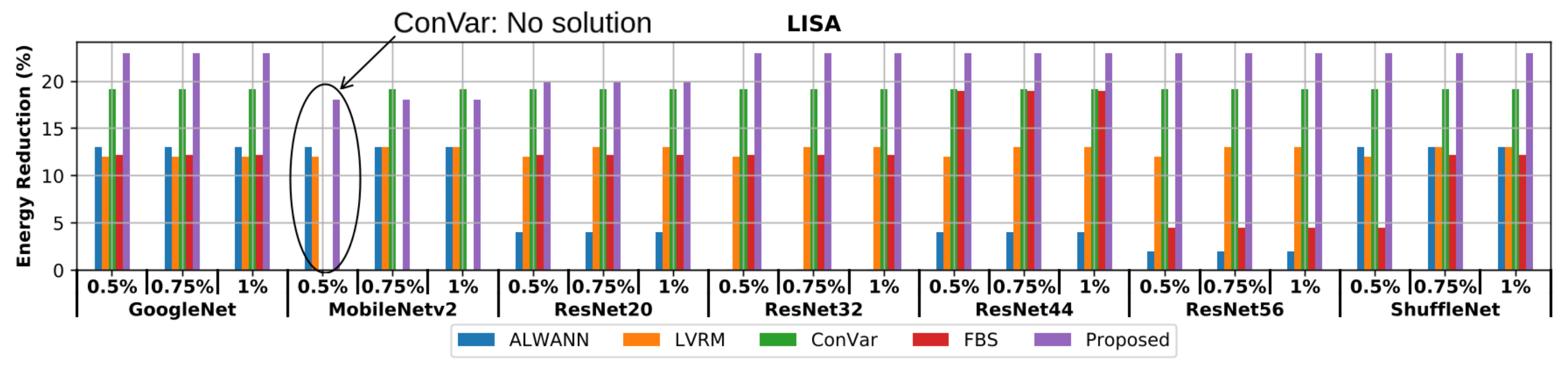}
    \caption{LISA normalized MAC operation energy savings. For the $0.5\%$ threshold for the MobileNetv2 case, ConVar~\cite{zervakis2021control} could not produce acceptable solutions, while ALWANN~\cite{mrazek2019alwann} resulted in nearly $0\%$ gains in energy for ResNet32.}
    \label{fig:lisa_results}
\end{figure*}

For each of the 4 datasets we considered, Fig.\ref{fig:cifar10_results}~-~\ref{fig:lisa_results} show the respective results for all NNs and all three different accuracy thresholds $0.5\%$, $0.75\%$ and $1\%$. Specifically, Fig.~\ref{fig:cifar10_results}~-~\ref{fig:lisa_results} 
report the energy reduction achieved by our method as well as the state of the art. Energy reduction is calculated w.r.t. the energy consumption of the exact design.
As mentioned in Section~\ref{sec:methods}, for ALWANN~\cite{mrazek2019alwann} we considered all the approximate multipliers from the library in~\cite{mrazek2017evoapproxsb} and show the results from the multiplier that yielded the highest energy gains for each accuracy threshold. Additionally, we evaluated FBS for $z\in[1,3]$ and included the results that yielded the highest energy gains.

Fig.~\ref{fig:cifar10_results} shows the energy savings for the CIFAR-10 dataset. 
Overall, across all $7$ NNs our approach achieves an average of $17.43\%$ in energy gains for all the considered accuracy thresholds when compared to the exact mode of operation. 
Specifically, our method sustains an $18\%$ energy reduction for the ResNet20 and ResNet32, $22\%$ for the ResNet44 and  ResNet56, and $12\%$ for the ShuffleNet. Some variation in energy gains is observed for the MobileNetv2 (from $11\%$ to $13\%$), and the greatest variation is observed for the GoogleNet, where our method achieves from $14\%$ up to $20\%$ energy gains. Furthermore, we observe that the energy reduction gains increase as the NNs become deeper with more layers (e.g., ResNet56). 
ALWANN~\cite{mrazek2019alwann} and LVRM~\cite{tasoulas2020weight} achieve on average $3.77\%$ and $11.57\%$ gain in energy respectively. 
The multiple modes of approximation that LVRM~\cite{tasoulas2020weight} introduces can adapt to various accuracy thresholds, and in this case manages to slightly surpass our proposed method's gains by $1\%$ for the $0.5\%$ threshold for MobileNetv2. The ConVar~\cite{zervakis2021control} method sustains a $19.2\%$ gain in energy, surpassing our proposed method's results by $1.77\%$, while FBS resulted in an average of $7.8\%$ in energy gains.


Fig.~\ref{fig:cifar100_results} shows the energy savings for the CIFAR-100 dataset. For this dataset, the achieved gains are the lowest we observed in our evaluation since it is a more challenging dataset~\cite{mrazek2020libraries}. 
However, our proposed method still attained an average gain of approximately $15.33\%$ across all NNs and accuracy thresholds.
The lowest gains on this dataset were observed for the ResNet20 ($8\%$) and for the MobileNetv2 ($12\%$). However, our method's maximum achievable gains in energy for this dataset are still high, reaching up to $22\%$ for the ResNet56 and $21\%$ for the ResNet32. For the Mobilenetv2 our method achieves similar energy gains to LVRM~\cite{tasoulas2020weight} (from $11\%$ to $13\%$), and for the ResNet20 LVRM~\cite{tasoulas2020weight} surpasses our method's gains by an average of $4.5\%$ for the three accuracy thresholds. Overall, LVRM~\cite{tasoulas2020weight} and ALWANN~\cite{mrazek2019alwann} achieve an average of $11.71\%$ and $2.88\%$ in energy gains respectively, i.e., $1.31$x and $5.32$x lower than our method.
Again, ALWANN~\cite{mrazek2019alwann} does not result in significant energy gains for this dataset, while LVRM~\cite{tasoulas2020weight} maintains energy gains above a minimum $11\%$, without however achieving energy gains above $13\%$ across all NNs.
In this dataset, ConVar~\cite{zervakis2021control} again reached a $19.2\%$ in energy gains, however not across all considered NNs. Specifically, the considered multiplier \emph{could not satisfy} neither of the $0.5\%$ and $0.75\%$ accuracy thresholds for ResNet44, yielding no results in these cases ending up in an average of $17.37\%$ in energy gains for the CIFAR-100 dataset. Similarly, FBS did not produce acceptable results for any of the accuracy thresholds for GoogleNet and ResNet20 and only satisfied the $1\%$ threshold for the MobileNetv2 and ResNet32 NNs, while also failing to satisfy the $0.5\%$ threshold for ResNet56, ending up with an average gain in energy of just $2.11\%$. This behavior validates our argument that the proposed filter-oriented error balancing method produces better results. 

Fig.~\ref{fig:gtsrb_results} depicts the energy savings for the GTSRB dataset. Our proposed method achieves an average of $18.71\%$ in energy savings across all NNs and accuracy thresholds, while LVRM~\cite{tasoulas2020weight} and ALWANN~\cite{mrazek2019alwann} achieve $11.81\%$ and $3.44\%$ respectively. The lowest observed value in the attained energy savings of our method is approximately $12\%$ for the MobileNetv2, while the highest is $23\%$ for GoogleNet for the $1\%$ threshold. Excluding MobileNetv2, our method shows energy gains of a minimum of $17\%$ for the rest of the NNs while respecting all accuracy requirements. Again, ConVar~\cite{zervakis2021control} \emph{could not produce an acceptable result} for the $0.5\%$ threshold for ShuffleNet, ending up to an average of $18.29\%$ in energy savings. FBS also failed to produce acceptable results, specifically for ShuffleNet and for all the considered thresholds, resulting in $9.98\%$ energy savings on average across the dataset.

Finally the corresponding results for the LISA dataset are shown in Fig.~\ref{fig:lisa_results}. For this dataset, our method achieves the highest energy gains observed throughout our evaluation experiments, being $21.86\%$ on average (across all NNs and thresholds). In this dataset, LVRM~\cite{tasoulas2020weight} and ALWANN~\cite{mrazek2019alwann} achieve $12.57\%$ and $7\%$ respectively.
Specifically, the minimum observed energy gains value of our proposed method was $18\%$ for MobileNetv2. Additionally, our method achieved $20\%$ in energy gains for ResNet20 and $23\%$ for the rest of the NNs for all the accuracy thresholds. For this dataset, the average gain in energy that ALWANN~\cite{mrazek2019alwann} achieved was doubled in comparison to the other datasets. For the MobileNetv2, ShuffleNet and GoogleNet NNs, the gains of LVRM~\cite{tasoulas2020weight} were similar and in some threshold cases lower than the ones of ALWANN~\cite{mrazek2019alwann}. For the $0.5\%$ for MobileNetv2, ConVar~\cite{zervakis2021control} \emph{again failed to meet the requirement} and it did not result in acceptable accuracy, ending up to an average of $18.29\%$ in energy savings for this dataset as well. Likewise, FBS did not produce any acceptable results for none of the considered thresholds for MobileNetv2, ending up with an average of $9.94\%$ in energy gains.

The NNs that we considered in our evaluation (shown in Fig.~\ref{fig:cifar10_results}~-~\ref{fig:lisa_results}) were trained on $4$ different datasets, and the methods we included in our comparison did not require retraining. Our method resulted in solution mappings that respected all considered accuracy thresholds for all NNs, yielding high gains in energy for every case. 
Our methodology achieves overall higher energy gains when compared to the corresponding reconfigurable weight-oriented method presented in LVRM~\cite{tasoulas2020weight}, surpassing it in some cases by as much as $12\%$. On average, our method achieved $18.33\%$, while ConVar~\cite{zervakis2021control} yielded $18.29\%$, LVRM~\cite{tasoulas2020weight} $11.9\%$, FBS $7.46\%$ and ALWANN~\cite{mrazek2019alwann} $4.27\%$ in terms of energy savings. ConVar~\cite{zervakis2021control} was the method that reached and maintained energy savings similar to our proposed method's, in some cases surpassing our results by up to $11\%$ (ResNet20 for CIFAR-100). However, 
ConVar~\cite{zervakis2021control} failed \emph{repeatedly} to satisfy the given accuracy thresholds as opposed to our technique that always satisfied the accuracy constraints.
By pushing the approximation more we were able to counteract smaller energy gains in some cases with greater gains in others, reaching average energy savings similar to ConVar~\cite{zervakis2021control}.
FBS also failed to produce acceptable solutions on multiple occasions, justifying our choice to only employ LDM on smaller sets of weight values in the final step of our mapping methodology.

\section{Conclusion}\label{sec:Conclusion}

In this work, we present an approximate multiplier that can be configured to generate positive, negative, or no error.
Our mathematical analysis demonstrates that by leveraging the known weight values to carefully set the modes of our approximate multiplier per weight, we can minimize the convolution error and thus attain high inference accuracy.
To achieve this, we propose a filter-oriented mapping methodology that aims to satisfy a given accuracy drop threshold while maximizing the applied approximation, targeting high energy efficiency.
Our extensive experimental evaluation, shows that our filter-oriented approximation with our positive/negative multiplier outperforms significantly the state of the art.
It is noteworthy that our proposed technique does not require DNN retraining.

\section*{Acknowledgement}
This work is supported in part by the German Research Foundation (DFG) through the project ``ACCROSS: Approximate Computing aCROss the System Stack''.

\newpage
\balance

\end{document}